\documentclass{aastex}
\usepackage{emulateapj5}

\usepackage{graphicx}

\shorttitle{Unstable disks at high redshift: Evidence for smooth accretion}

\shortauthors{Bournaud \& Elmegreen}

\begin{document}

\title{Unstable disks at high redshift: Evidence for smooth accretion in galaxy formation}

\author{Fr\'ed\'eric Bournaud}
\affil{CEA, IRFU, SAp, 91191 Gif-sur-Yvette, France}
\email{frederic.bournaud@cea.fr}

\author{Bruce G. Elmegreen}
\affil{IBM Research Division, T.J. Watson Research Center, 1101
Kitchawan Road, Yorktown Heights, NY 10598, USA} \email{bge@us.ibm.com}

\begin{abstract}
Galaxies above redshift 1 can be very clumpy, with irregular
morphologies dominated by star complexes as large as 2 kpc and as
massive as a few $\times10^8$ or $10^9$~M$_{\sun}$. Their co-moving
densities and rapid evolution suggest that most present-day spirals
could have formed through a clumpy phase. The clumps may form by
gravitational instabilities in gas-rich turbulent disks; they do not
appear to be separate galaxies merging together. We show here that the
formation of the observed clumps requires initial disks of gas and
stars with almost no stabilizing bulge or stellar halo. This cannot be
achieved in models where disk galaxies grow by mergers. Mergers tend to make stellar spheroids even when the gas fraction is high, and then the disk is too stable to make giant clumps. The morphology of
high-redshift galaxies thus suggests that inner disks assemble mostly
by smooth gas accretion, either from cosmological flows or from the
outer disk during a grazing interaction.
\end{abstract}


\section{Introduction}

In the standard $\Lambda$-CDM cosmology \citep{blumenthal84}, galaxies
assemble through hierarchical merging \citep{davis85} and then evolve
under a combination of internal and environmental processes. Most of a
galaxy's mass comes from mergers with much smaller objects and major
mergers are less frequent \citep{delucia06, genel08}. Major and minor
mergers tend to transform disks into spheroids and ellipticals
\citep{naab07,bournaud07}, but if the gas fraction is high enough, then
massive and extended disk structures can persist
\citep{robertson06,springel05,robertson-bullock08}. Therefore it may be
possible, within the hierarchical framework, to explain the formation
of disk galaxies with morphologies and kinematics like those observed
locally \citep{governato07}. It has recently been suggested, however,
that a large part of the galaxy mass could instead come from diffuse
gas accretion, in particular along cold flows
\citep{dekel09a,ocvirk08,keres08}. Thus, the build-up of high-redshift
galaxies may result from two processes, one hierarchical and the other
somewhat smooth. This {\it Letter} considers the resolved properties of
these galaxies to suggest that the smooth process dominates in a high
fraction of cases.

Galaxies are increasingly clumpy with redshift \citep{conselice05}.
These clumps are not just features in otherwise normal spirals and
ellipticals \citep{cowie96,vdb97}. Most high-redshift galaxies do not
have spirals in restframe blue and uv bands, nor do they have bulges or
exponential profiles. Usually, a large fraction of their optical light
(up to 50\%) and luminous mass (up to 30\%) is confined to a few
kpc-size clumps \citep[][hereafter EE05]{EE05}. Even the inter-clump
light does not follow a spiral- or exponential-like profile
\citep{EEVFF05}. Highly aligned clumps have been called {\em chain}
galaxies \citep{cowie96}, while rounder systems have been
called {\em clump-clusters} \citep{EEH04}. Both types could be
progenitors of modern spiral disks, viewed with different orientations
(Elmegreen et al. 2005; Bournaud, Elmegreen \& Elmegreen 2007,
hereafter BEE07). If the clumps are massive enough, then they spiral to the center to make a bulge (\citealt{noguchi}; \citealt{immeli};
Elmegreen, Bournaud \& Elmegreen 2008).

Clumpy galaxies are so frequent at $1 \leq z \leq 5$, and the clumps
evolve so quickly, that most present day spirals could have had a
clumpy phase in their past. Their comoving space density of the UDF is
$\sim10^{-3}$ Mpc$^{-3}$ between $z=1$ and 4, comparable to the density
of spirals in the same redshift range and with the same absolute
$z_{850}$ magnitude \citep{elmegreen07}. Moreover, the lifetime of the
clumpy phase is short, $\sim0.5$ Gyr, according to simulations (BEE07).
If we consider the ratio of the lifetime of the clumpy phase to the
Hubble time as a measure of the fraction of galaxies in that phase at
any one time, and multiply the inverse of this ratio by the observed
space density of the clumpy systems, then we get a total space density
for galaxies that ever went through the clumpy phase. This total is
comparable to the space density of modern spirals.

Several observations suggest that clumps in most chain and
clump-cluster galaxies formed inside their disks rather than entered
from outside in a merger. First, the distribution of the ratio of axes
for the combined population is approximately flat, suggesting that most
of the clumpy types are disks viewed in random orientations (EE05).
Second, the clumps in chains are highly confined to the average
midplanes, which makes external capture unlikely \citep{elmegreen06}.
Third, the masses and sizes of the clumps are consistent with their
formation by gravitational instabilities for the observed velocity
dispersion \citep{FS06,genzel08} and a gas column density comparable to
the total in today's inner spiral disks  \citep{E09}. Fourth, the
height of a clump is comparable to the disk half-thickness, suggesting
that both are determined by the gravitational scale length
\citep{elmegreen06}. Fifth, the distribution of relative clump position
in a UDF chain is the same as for edge-on clump clusters (Elmegreen
2009). Sixth, the largest clumps have similar masses and ages, unlike
the expectation for random capture \citep{E09}. The shape of clumps is
another clue to their {\it in situ} origin: they are not elongated like
spiral arms, so they have to form quickly in a highly unstable disk
(Toomre $Q\le1$). Red interclump colors and a monotonic rise in the
rotation curve for one studied case \citep{bournaud08} also suggest the
clumps are part of a disk.

Here we show that giant star-forming clumps require high turbulent
speeds and a dense disk with few stars in a spheroid. Simulations of
galaxy mergers predict something different: that a large fraction of
the stars should end up in a spheroid. Such a spheroid stabilizes the
disk and prevents the observed clumps from forming.

\section{Requirements for giant clump instabilities in primordial gas disks}

\subsection{High velocity dispersions}

Typical attributes of UDF clumpy galaxies are a stellar mass $M_* =
6\times 10^{10}$~M$_{\sun}$ and a disk radius $R = 9$~kpc (EE05). The
gas mass fraction $f_{\mathrm g}$ (gas to total baryon ratio)
is not known observationally. An estimate of $f_{\mathrm g} \simeq 50\%$ was made by \citet{daddi08} in BzK-selected galaxies. We apply this fraction to clumpy galaxies because (i) Daddi et al.'s galaxies are
clumpy in the HST images, and (ii) other BzK-selected galaxies have
massive clumps like the UDF galaxies \citep{FS06,genzel08}.
\citet{bouche07} also support gas fractions around 50\%. Much larger
gas fractions would exceed the dynamical mass from circular velocities \citep{bournaud08, daddi08}. Lower gas fractions would make the clumps too big for the observed velocity dispersion. Thus, we assume $f_{\mathrm g} = 50\%$ inside the stellar disk radius, and a gas mass $M_{\mathrm g} = M_* = 6\times 10^{10}$~M$_{\sun}$.

The Jeans length for gravitational instabilities is $\lambda_{\mathrm J
} = \sigma^2/\left(\pi G \Sigma\right)$. For somewhat uniform disks,
$\Sigma\sim M_{\mathrm g}/\left(\pi R^2\right)$, which gives,
\begin{equation}
\sigma^2\sim\frac{\lambda_{\mathrm J } G M_*}{R^2}.
\end{equation}
We consider clumps of size $\lambda_J\sim500$ pc or larger, so $\sigma
\simeq 50$~km~s$^{-1}$ for the gas that forms giant clumps. Lower
velocity dispersions would give smaller clumps. This result is
consistent with H$\alpha$ observations that suggest a high turbulent
speed \citep{FS06,genzel08}, and with observations of moderately thick
disks \citep{elmegreen06}.
\medskip

\subsection{Dense disks with low-mass stellar spheroids}
\subsubsection{Analytical constraints on $Q$}

Clumps form by gravitational instabilities if $Q=\sigma\kappa/\left(\pi
G\Sigma\right)\le1$.  We consider first this requirement on $Q$ for a
model in which (1) M$_{\mathrm g}=6\times 10^{10}$~M$_{\sun}$ in a disk
of radius $R=9$~kpc with a turbulent speed $\sigma=50$~km~s$^{-1}$; (2)
the total stellar mass of M$_{*}=6\times 10^{10}\;M_{\sun}$ inside $R$
is partly in a disk of mass M$_{*,D}$, partly in a central bulge of
mass M$_{*,B}$ and radius 1~kpc, and partly in a stellar spheroid with
M$_{*,H}$=1-M$_{*,D}$-M$_{*,B}$; (3) there is always a dark halo with a mass
$f_{\mathrm D} \times (M_{\mathrm g}+M_{*})$ inside $R$. We choose
$f_{\mathrm D}$=0.5, which means that 2/3 of the mass inside $R$ is
baryonic and 1/3 is dark. High redshift kinematic observations (Daddi
et al. 2008, Bournaud et al. 2008) are consistent with this fraction.
Modern disk galaxies should be about the same because late infall of
dark matter and baryons typically follow each other (Semelin \& Combes
2005). Locally, the dark-to-disk ratio $f_D$ is on average 0.6-0.7, and
most generally in the 0.3-1.0 range (e.g., Persic \& Salluci 1990); our
$f_D=0.5$ choice is thus realistic. We further discuss the influence of
this parameter in Section~3.

We compute $Q$ for the combined gas and stellar disks, assuming the
stars have about the same velocity dispersion as the gas because both
are heated by gravitational instabilities and clump interactions. Then
\citep{wang-silk94}:
\begin{equation}
{Q(r)} \sim \left(\frac{\pi G \Sigma_{\mathrm gas}}{\sigma_{\mathrm
gas} \kappa(r)} + \frac{3.36 G \Sigma_{\mathrm star}}{\sigma_{\mathrm
star} \kappa(r)}\right)^{-1}.
\end{equation}
When $\sigma$ is the same for gas and stars, $Q\simeq
\sigma\kappa/\left(\pi G \left[\Sigma_{\mathrm gas}+\Sigma_{\mathrm
star}\right]\right)$. Profiles of $Q(r)$ are shown on Figure~1. When
all the stars are in the disk, $\Sigma$ is high enough and $\kappa$ is
low enough that $Q<1$ over a large part of the disk. When less than
80\% of the stars are in the disk and more than 20\% are in a halo or
bulge, $Q>1$ over the whole disk. Then spiral arms can form with star-forming clumps inside, but isolated and round clumps become less likely.

\subsubsection{Numerical Simulations}

To check these simple $Q$ estimates, we ran models of galaxies with
various ratios of disk:bulge:spheroid stellar mass and determined when giant clumps appeared. The mass and size parameters are as
in the calculations above, and we start with $\sigma$=50~km~s$^{-1}$
for gas and stars. The simulations were run with higher resolution than
in BEE07: $3 \times 10^6$ particles each for the stars, gas, and dark
matter, and a spatial resolution of 30~pc over the whole disk. The
Jeans length is more than ten times larger than the resolution in our
initial set-up, so numerical fragmentation is avoided \citep{truelove}.

Figure~2 shows the mass distribution of gas and stars at the most
clumpy instant for several models. When 100\% percent of the stellar
mass is initially in the disk, dense round kpc-sized clumps form as in
the observed high-redshift galaxies. When 80\% of the stars are in the
disk and 20\% are in a bulge or extended stellar halo, the disk is
still somewhat clumpy, but the clumps are not isolated, round and
gravitationally bound -- they are bright spots in shearing spiral arms.
When 60\% or less of the stars lie in the disk and the rest are in a
bulge or halo, the disk forms mostly spiral arms (small, low-mass
clumps can still form, but only when $\sigma$ is small).

These models confirm the earlier calculations based on $Q$ alone. The formation of giant clumps requires $\sigma$ and $\Sigma$ to be high and $\kappa$ to be low (Eq.~2). Stars in the rotating disk contribute to $\Sigma$ and help drive the clump-forming instability, while stars in the bulge or spheroid increase $\kappa$ and stabilize the disk. Thus the formation of giant clumps in galaxies with masses and sizes typical of the UDF clump clusters requires that almost all of the stars lie in a rotating disk together with the gas when clump formation starts. If this constraint is not satisfied, then $Q>1$ and massive clumps do not readily form. If turbulent dissipation decreases $\sigma$ enough to form clumps when the stellar disk mass is relatively small, then the resulting clumps will have low masses, unlike the observed clumps. Such small clumps are seen in the outer parts of the 60:30:10 and 60:10:30 models in Fig.~2. The large observed sizes of clumps at high redshift imply that $\sigma$ is high, and values of $\sigma$ higher than the one conservatively assumed above would make the formation of clumps even harder to achieve if a stellar spheroid is present, by increasing the $Q$ parameter.

\section{Clumpy galaxies in the hierarchical context}
The clumpy galaxies in the UDF are not obviously mergers, although
outer tidal tails and other merger debris could be too faint to see.
They are also not former mergers because most of their stars should be confined to a disk in order to get the giant clumps.  An important
characteristic of disks built by galaxy mergers is that they are
embedded in stellar spheroids. For instance, the model spiral galaxy
formed in a major merger by \citet{springel05} has only $\simeq$ 25\%
of its stellar mass in the rotating disk, while $\simeq$ 50\% is in a
massive, kpc-sized bulge and $\simeq$ 25\% is in a more extended
diffuse stellar halo. Even mergers starting as pure gas rarely end-up
with more than 50\% of the stellar mass in the disk:
\citet{robertson06} find that for gas fractions of 0.4 or 0.6,
equal-mass mergers end up with only 10\% or less of the stars in the
disk, the rest being in a bulge and a stellar halo; even initial gas
fractions of 80\% produce disks containing only 40-45\% of the stellar
mass \citep[models DC, EC, FC in][]{robertson06}. In the case of minor
mergers, their models DCm, ECm, FCm indicate final disk fractions of
60-75\% after a single 8:1 merger, for gas fractions ranging from 0.4
to 0.8. A larger study by \citep{hopkins08} also finds that the
fraction of stars in the disk after a single 8:1 minor merger is 70\%
to 85\%, depending on the gas fraction and other parameters (e.g.,
their Fig.~12); after a 2:1 merger, the mass in spheroids is equal to
or larger than the stellar disk mass. A 1:1 merger ends-up with less
than one third of the stellar mass in the rotating disk component.

If we consider a typical clumpy galaxy in the UDF and assume that, over
the last doubling of its mass, mergers were the dominant growth
process, then that growth may have occurred, for example, by a 1:1
major merger or by six successive 8:1 minor mergers. For gas fractions
of 50\% in the progenitor galaxies \citep[likely for
$z\sim2$,][]{daddi08}, such mergers will leave only 20\% or less of the
stellar mass in the rotating disk, the vast majority being in the bulge
and halo. Even higher gas fractions will not allow more than $\sim
50$\% of the stellar mass in the disk. If we assume that the last mass
doubling event was half done by mergers and half by smooth accretion of
cold gas, then this requires one 2:1 merger, or 3 to 4 mergers of mass
ratio 8:1. These will leave only 40-50\% of the stellar mass in the
disk, given the various results mentioned above. If we consider mass
assembly over longer periods, then the effect of mergers on the bulge
and stellar halo will be even more dramatic.
\smallskip

The ubiquity of giant clumps in $z\sim 2$ disk galaxies can be
explained most easily if only a small fraction of the stars lie in a bulge or halo before the clumps formed. We found that this fraction should be $<20$\%, assuming a dark to baryonic ratio $f_D$ of 50\% inside the
optical radius. This ratio $f_D$ should not be much lower in the
progenitors of present day spirals. For a very low $f_D$ of 25\% (80\%
of the mass inside $R$ is baryonic), a slightly more massive stellar
spheroid could be present without preventing the formation of large
clumps. Still, the fraction of baryons that can be in a bulge or halo remains limited to $\leq 30$\% even in this extreme case (to keep the total dark+stellar spheroid constant). This would allow a relatively higher contribution from mergers in the formation of these galaxies, but it still implies that most of the mass has to enter in a smooth accretion process instead of through major and minor mergers.

The key mechanism is unlikely to be distant interactions that
de-stabilize the disk and provoke internal clump formation.
Interactions can destabilize a disk (e.g., di~Matteo et al. 2008) when
the disk was marginally stable, but not when $Q$ was originally high
because of a massive spheroid. Furthermore, an interaction-induced
process would lead to centrally-concentrated star formation because of
the associated angular momentum redistribution. This would contradict
observations often showing the star-forming clumps at large radii or in ring-like
structures in bulge-free disks (Elmegreen \& Elmegreen 2006b; Genzel et
al. 2008; Elmegreen et al. 2009).

Gas physics and thermal cooling can affect the long-term evolution of
clumps, as studied with various hydrodynamic codes by Immeli et al.
(2004), Debasttista et al. (2006) and Tasker \& Bryan (2008). However,
in both our simulations and our analytical calculations, thermal
cooling is not a critical issue because the turbulent motions in the
gas greatly exceed the thermal motions on kpc scales where the
instability operates. For our simulations, the turbulent speed is set
to a realistic value considering the observed clump sizes, disk
thicknesses, and gas velocity dispersions. The medium in which we form
clumps thus has realistic properties. Only the small-scale properties
of the clumps, such as core formation, molecule formation, and dense
cluster formation, all of which are unobserved so far, can be affected
by thermal cooling. In a supersonic medium, thermal cooling rates
affect the thickness of the shock fronts, but not the overall
dissipation rate of the turbulent motions. In this regard, we note that
large-scale simulations of supersonically turbulent gas and clump
formation require numerical methods that can handle high velocity
dispersions in a thermally cold interstellar medium.
\medskip

If most of the mass assembly of these galaxies came from major and
minor mergers, then the disk density would be too low and the shear rate
would be too high. Clumps could not form with the masses and sizes
typically observed at high redshift. For instance, the merger-produced
disk in the model by \citet{robertson-bullock08} is massive and
extended, but it only grows spiral arms without giant clumps. The
assembly of high-reshift galaxies, with frequent clumpy morphologies, is unlikely to be mostly driven by hierarchical merging of smaller galaxies. Even if minor and major mergers were responsible for only half of the growth, the spheroids would be too massive for disk
clumps. The formation of giant clumps points to massive, highly turbulent disks that have relatively small bulges and stellar haloes. This requires that the dominant process of mass assembly be some smooth accretion of cold and diffuse gas. This conclusion is consistent with the recent picture in which young thick disks form by cold flows (Dekel et al. 2009a; Keres et al. 2008) and
other types of diffuse gas accretion \citep{semelin}, bulges form by
internal, clump-driven evolution \citep{EBE08,genzel08}, and the thin
disk forms later by further smooth accretion \citep[e.g.,][]{BC02}. Our
result does not imply that mergers do not occur at high redshift, but
only that they cannot be the main mechanism for disk assembly.

\section{Conclusion}

The ubiquity of giant clumps in high-redshift disk galaxies constrains
the mass distribution and therefore assembly process. Most of the
stellar and gas mass should be in a disk, rather than in a bulge or
spheroidal halo, in order to get the clump masses, sizes, and morphologies correct. This implies that the assembly is mostly smooth, with only a small fraction through minor and major mergers. Bulges then form as an aftermath of clump evolution. An exception
might occur for S0 galaxies, which have much larger stellar spheroids;
mergers could have played a more important role in their formation and structure.

Cosmological models including warm dark matter could lead to the required smooth gas accretion because warm dark matter has relatively little substructure. The model by Heller, Shlosman \& 
Athanassoula (2007) had a monolithic collapse of gas inside a single halo without hierarchical merging: the gas flow 
was smooth, and the disk formed through a clumpy phase. 
Cold dark matter models may also produce smooth gas accretion and satisfy our constraint on the disk mass fraction (e.g., Dekel et al. 2009a; Semelin \& Combes 2005; Keres et al. 2008).
Recent models by \citet*{dekel09b} and \citet{agertz} suggest that cold flows could maybe lead to the formation of clumpy galaxies.

Present-day spiral galaxies have apparently evolved through a clumpy
phase, and this observation may prove to be a key factor in
understanding galaxy assembly and the nature of dark matter.

\acknowledgments
We are grateful to Debra Meloy Elmegreen for useful discussions, and to an anonymous referee for constructive comments. Numerical simulations were carried out on the NEC-SX8R vector computer at CEA/CCRT.

{}

\begin{figure}
\centering
\includegraphics[width=4in]{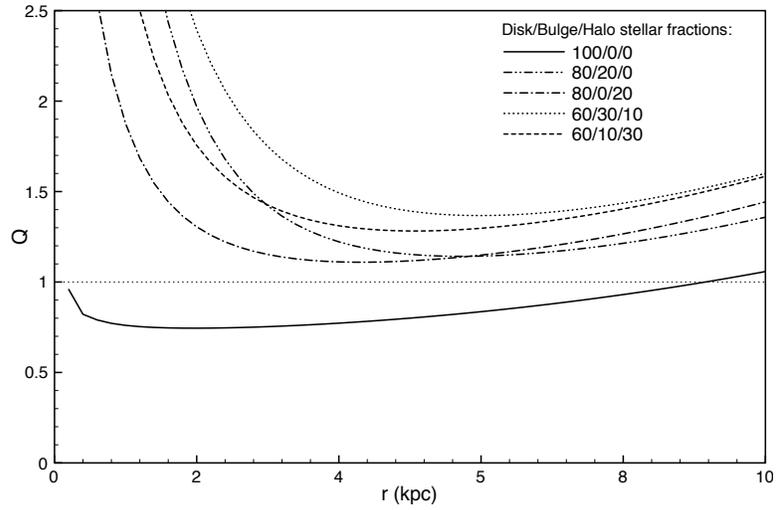}
\caption{Toomre $Q$ parameter for the gaseous and stellar disks, with
several assumptions about the stellar mass distribution. The legend
indicates the proportion of disk:halo:bulge stars. We assume a uniform
surface density for the disk, a Hernquist profile with a scale-length
of 600~pc for the bulge, and the same profile with a 6~kpc scale-length for the stellar halo. There is always a dark matter halo, with a core Burkert profile (a cuspy profile would somewhat
increase $Q$).}\label{fig:1}
\end{figure}

\begin{figure}
\centering
\includegraphics[width=6in]{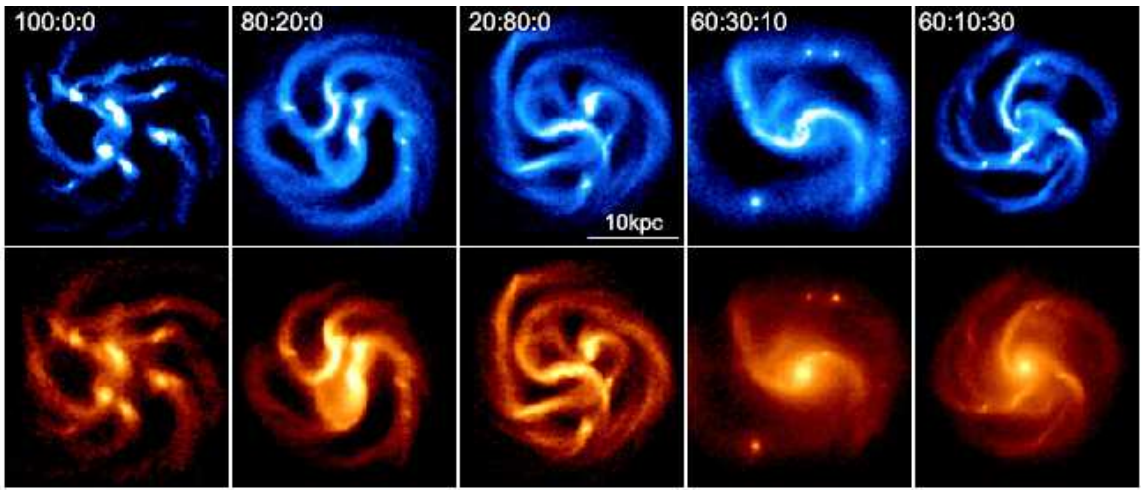}
\caption{Surface density maps for the gas (top) and stars (bottom), at
the most "clumpy" instant seen face-on. The {\it stellar} mass distribution in the disk:bulge:halo components is varied as in Fig.~1; there is always a gas disk and a dark matter halo. All snapshots are shown with the same logscale colorbar.}\label{fig:2}
\end{figure}

\end{document}